# Substrates modulate charge-reorganization allosteric effects in protein-protein association

Shirsendu Ghosh,[#] Koyel Banerjee-Ghosh,[#] Dorit Levy, Inbal Riven,

Ron Naaman,* Gilad Haran*

Department of Chemical and Biological Physics, Weizmann Institute,

Rehovot 76100, Israel

[#]The first two authors contributed equally.

ABSTRACT

Protein function may be modulated by an event occurring far away from the functional site, a phenomenon termed allostery. While classically allostery involves conformational changes, we recently observed that charge redistribution within an antibody can also lead to an allosteric effect, modulating the kinetics of binding to target antigen. In the present study, we study the association of a poly-histidine tagged enzyme (phosphoglycerate kinase, PGK) to surface-immobilized anti-His antibodies, finding a significant Charge-Reorganization Allostery (CRA) effect. We further observe that PGK's negatively charged nucleotide substrates modulate CRA substantially, even though they bind far away from the His-tag-antibody interaction interface. In particular, binding of ATP reduces CRA by more than 50%. The results indicate that CRA may be affected by charged substrates bound to a protein and provide further insight into the role of charge redistribution in protein function.

**TOC GRAPHIC**

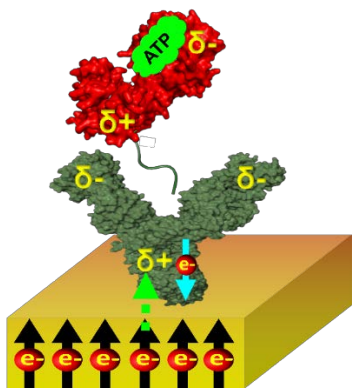

An important aspect of protein function and stability is the relative role of local versus non-local, long-range, interactions. Allostery, a prime example for a non-local effect, is defined as the modulation of the function of a protein by a perturbation that takes place at a site far away from the active site. This phenomenon is essential in biology, as proteins use allostery to control their biological activity.[1] The classical model of allostery invokes a conformational change that follows the binding of an allosteric modulator to a protein.[2,3,4] However, in our recently published study we showed that allostery can also be driven by a reorganization of charges within a protein that results from a modulation of its polarizability.[5] We demonstrated this charge-reorganization allostery (CRA) by studying the effect of charge injection on antibody-antigen association kinetics. In the present study, we discover that a small charged ligand that binds to the antigen may significantly modulate the CRA effect. This new concept of charge reorganization in proteins is consistent with the recent discovery of long range electron conduction in proteins[6] and the finding of long range modulation of electric field in proteins.[7]

We study here a polyhistidine (His)-tagged version of the enzyme phosphoglycerate kinase (PGK) from *Saccharomyces cerevisiae,* Baker's yeast, as an antigen that binds from solution to a surface-adsorbed anti-His antibody. PGK is an essential enzyme in glycolysis, which generates ADP and 1, 3-di-phosphoglycerate from ATP and 3-phosphoglycerate, or vice versa. Nucleotide binding to the enzyme requires a concomitant binding of a magnesium ion.[8] In our experiment, the His-tag is attached at the C-terminus of PGK and we study the kinetics of its interaction with an anti-His antibody attached to a magnetized metal surface, which serves as an electron source.

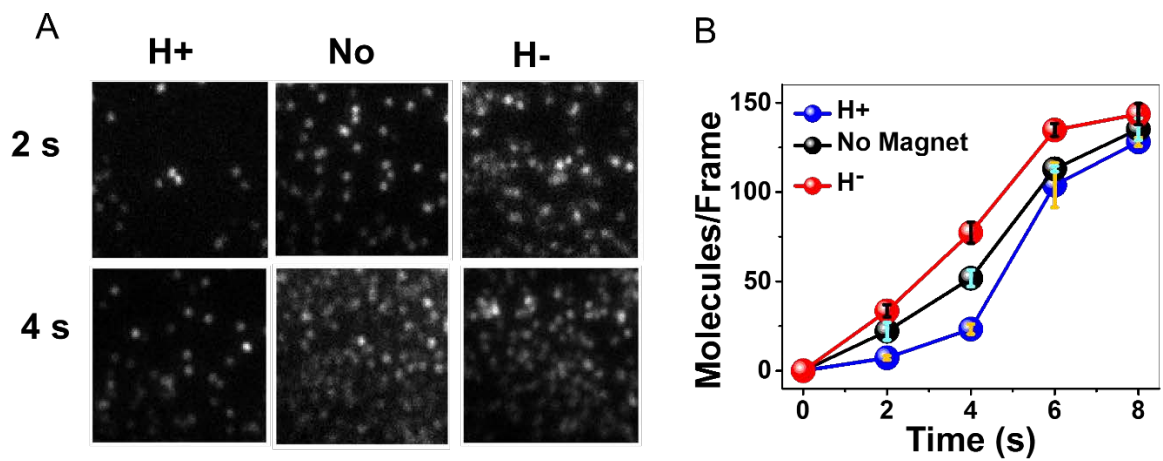

**Figure 1.** CRA in the binding of His-tagged PGK to an anti-His antibody. (A) Fluorescence images of individual complexes formed between anti-His antibodies adsorbed on a magnetized surface and fluorescently labelled His-tagged PGK molecules. The images are taken following two different interaction periods and with the North pole of the magnetic field pointing either UP (H+) or DOWN (H−) and also in absence of a magnet (No). (B) Antigen-antibody association kinetics under the two magnetic orientations and without the magnet. Error bars represent standard errors of mean.

Specifically, the antibody is adsorbed on a gold (2 nm)-coated Ni (120 nm) surface using dithiobis [succinimidyl] propionate (DSP) as a linker. The rate of the antigen-antibody association is measured with the substrate magnetized either with its North magnetic pole pointing UP (H+) or DOWN (H-), or with a non-magnetized substrate. Fluorescently labeled His-tagged PGK molecules are added to the solution on top of the magnetized substrate to initiate binding. At specific time intervals, following the initiation of the association reaction, the substrate is taken out of solution and immediately rinsed. The number of bound antigen molecules is counted using a fluorescent microscope. From Figure 1A and 1B, it is revealed that the association reaction is faster when the surface is magnetized with the North pole of magnet pointing down (H-). As a control, the antigen-antibody interaction is also studied with a non-magnetized substrate, which as expected, gives a result that is essentially the average of the two other measurements (Figure 1B).

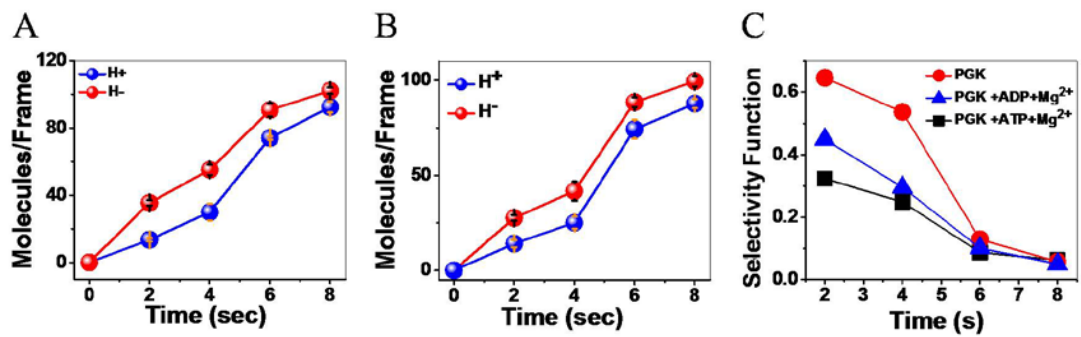

**Figure 2.** Substrate binding to PGK modulates the CRA effect. (A) Antigen-antibody association kinetics in the presence of Mg-ADP. (B) Antigen-antibody association kinetics in the presence of Mg-ATP. (C) Selectivity function of association kinetics with and without substrates. Error bars represent standard errors of mean.

These results are similar to our previously reported results, where the antigen used was a His-tagged version of a different protein, ClpB, and can be explained with the same CRA mechanism.[5] Briefly, when a protein molecule approaches the antibody, it induces charge polarization that results in charge reorganization within the antibody. This charge reorganization depends on the polarizability of the antibody, which is enhanced if electrons can move either from it to the metallic surface or vice versa. Since it is now well established that for chiral molecules charge polarization is accompanied by spin polarization,[9] facile electron transfer between the antibody and the metallic surface depends on the spin orientation of the electrons in the ferromagnetic substrate, which is modulated by the direction of the magnetic field. This effect is referred to as the chiral induced spin selectivity (CISS) and serves as an effective valve for the flow of electrons from the surface into the adsorbed protein molecules.[10] It can therefore lead to either acceleration or retardation of antibody-antigen association kinetics, depending on the orientation of the substrate's magnetization.

In the current case, as a PGK molecule approaches the antibody with its positively charged C-terminus (as observed from a calculation of the dipole moment of the molecule), it polarizes the antibody, inducing a negative charge at the interaction site with the antigen and a positive charge near the surface. Hence, there is an electric potential difference that induces electron flow from the metallic surface into the antibody. However, the rate of this flow depends on the surface magnetization, as discussed above.

We investigated the effect of the binding of negatively charged PGK substrates, Mg-ADP and Mg-ATP, on the CRA effect. As shown in Fig. 2A, the CRA effect with ADP is smaller than that shown in Fig. 1. ATP suppresses CRA even more, almost eliminating it (Fig. 2B). To show this more quantitatively, we compute the selectivity function $(N_{H-} - N_{H+})/(N_{H-} + N_{H+})$, with $N_{H\pm}$ the number of bound PGK molecules with the magnetic field either pointing up or down. Fig. 2C shows the selectivity functions for the three measurements. Since CRA affects the kinetics of protein-protein association, the selectivity is largest at the earliest time point and decays as time progresses. However, the selectivity is clearly reduced in the presence of ADP and ATP. In fact, ATP reduces the selectivity to less than half of its value without the substrates.

To interpret these results, we turn to the cartoons in Fig. 3. As already noted, in the absence of substrates the approach of the antigen leads to polarization of the antibody (Fig. 3A-B), which in turn leads to charge flow to or from the surface, further enhancing the CRA effect and increasing the rate of the interaction between the proteins. (For a detailed discussion of the role of the electron-flow valve induced by the CISS effect see Supporting Figure 1.) When a negatively charged substrate, such as ATP, binds to PGK (Fig. 3C), there are two potential mechanisms that may operate in parallel and lead to a change in the polarization of the PGK molecule.

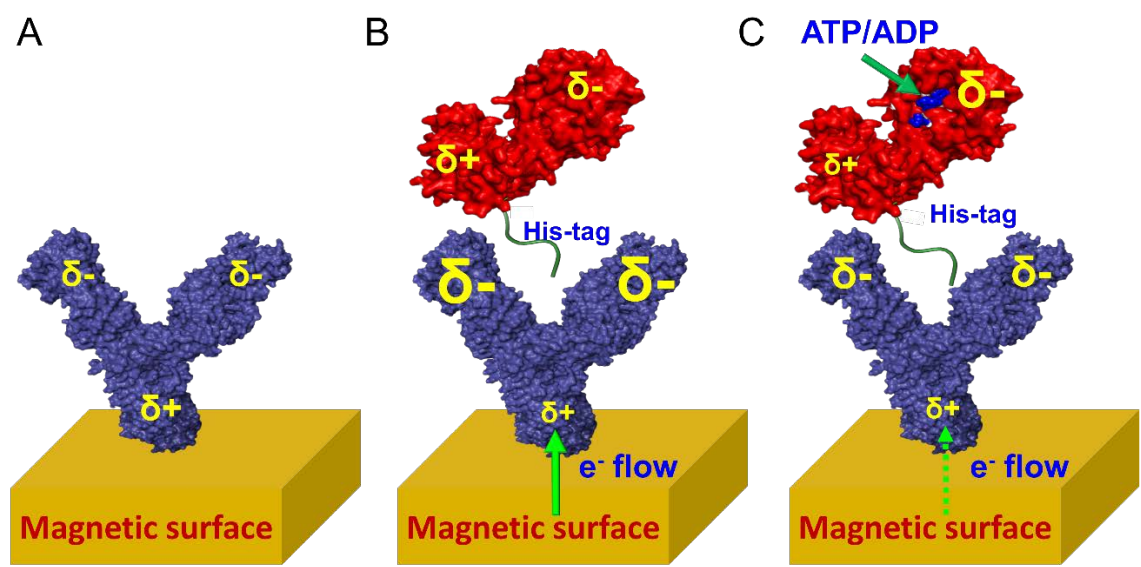

Figure 3. Representation of the events accompanying antigen-antibody interaction. (A) An antibody molecule attached to the surface has a certain distribution of charges, depicted here schematically. (B) the approach of a PGK molecule polarizes the antibody, leading to accumulation of more negative charge close to the binding site, and accordingly to a flow of electrons from the surface into the molecule. (C) When a negatively charged nucleotide is bound to PGK, the protein's dipole is reduced in size, and hence it does not polarize the antibody to the same extent as in B, and the electron flow from the surface is slower. The relative size of each δ sign indicates the amount of charge accumulated at a given electric pole.

First, the binding of the ATP moiety may lead to subtle changes in the positions of charged groups within the protein, driven by reorientations of residues and their side chains. Sato et al.[11] discussed this effect, which they termed "dielectric allostery", using molecular dynamics simulations of the binding of ATP to myosin. In the case of PGK, this would lead to a reduction of the overall dipole moment of the protein, therefore decreasing the polarization of the antibody by the approaching antigen, and in turn reducing the effect of charge flow from the surface on antibody-antigen association. In addition to this scenario, which is based on the orientational polarizability of PGK, there is another potential mechanism, based on the electronic polarizability of the protein. The binding of ATP may lead to electron transfer from the charged molecule to the protein. This electron transfer event might involve a full electron or only a partial charge, but in any case, it would also affect the overall dipole moment of the protein and thereby modulate its influence on the antibody as it approaches it, with a similar overall result. Therefore, both scenarios discussed here would lead to a modulation of the dipole moment of PGK and in turn to a decrease in the polarization of the antibody and a weaker effect on the association rate due to charge flow from the surface (Fig. 3C), as demonstrated experimentally in Fig. 2.

This study of antigen-antibody association kinetics, in which the antibody molecules are attached to a ferromagnetic substrate and the antigen is the His-tagged version of the protein PGK, confirms the CRA mechanism reported recently.[5] We further find here that the presence of negatively charged substrate, bound to the antigen molecules, reduces the polarization of the antibody upon binding and in turn reduces the amount of charge transferred from the metallic surface. These observations indicate the presence of charge reorganization in the proteins studied here over long distances. Further, the binding of small charged molecules to one of the proteins induces a CRA effect that significantly modulates their association, even though these small molecules bind at a site that is far away from their interaction region.

In recent years there is a growing interest in the effect of electric fields on protein structure and function,[12,13,14] including their role in enzyme catalysis.[15,16] The CRA presents a novel mechanism by which electric field can have a long range effect. While this work has probed the role of CRA in the interaction of two proteins, future studies will investigate the effect of CRA on other protein functions, such as enzymatic activity.

**Acknowledgements**

GH and RN acknowledge partial support from the Israel Science Foundation. RN acknowledges the support of the MINERVA foundation. We thank Peter Hamm for interesting discussions on the CRA effect.

# Supporting Information

# Substrates modulate charge-reorganization allosteric effects in protein-protein association

Shirsendu Ghosh,[#] Koyel Banerjee-Ghosh,[#] Dorit Levy, Inbal Riven,
Ron Naaman,* Gilad Haran*
Department of Chemical and Biological Physics, Weizmann Institute,
Rehovot 76100, Israel

**Experimental details**

*Expression, purification and labelling of PGK*

The DNA for phosphoglycerate kinase (PGK) fused to a C-terminal 6xHis tag was cloned into a pET28b vector. For site-specific labeling of PGK, the wild-type cysteine at position 97 was replaced by a serine, and a new cysteine residue was incorporated using site-directed mutagenesis, resulting in the C97S S290C PGK mutant.

The C97S S290C PGK plasmid was transformed into *E. Coli* BL21 pLysS (DE3) cells (Invitrogen), and grown in LB media at 37 °C, up to an optical density of 0.8. Protein expression was induced by the addition of 1mM IPTG, and cells were then incubated at 25 °C for overnight. Following expression, bacteria were harvested and the proteins were purified on a Ni-NTA resin, according to the manufacturer's instructions (GE Healthcare). Protein was kept until used at -80 °C in storage buffer (20 mM sodium phosphate, 1 mM TCEP, pH 6.8).

PGK was tagged after buffer exchange into labeling buffer (50 mM Tris, 25mM KCl, pH 7.1), using a desalting column (Sephadex G25, GE Healthcare). Protein was then mixed with Alexa 647 C2 maleimide (Invitrogen) at a 1:1.5 protein-to-dye ratio for 2h at RT. Labeled molecules were then separated from excess dye using a desalting column.

*Kinetic study of the interaction between His-tagged PGK and anti-His antibodies*

Ultra-LEAF™ Purified anti-His antibody was attached to the gold coated magnetic surface using dithiobis[succinimidyl] propionate (DSP) as a linker. Surface preparation and immobilization of the antibody were done as in our previous work.[1] A DSP monolayer was formed on the gold surfaces by incubating them in a solution of DSP in DMSO (4 mg/ml) for 30 mins. After that, the surfaces were rinsed with DMSO and water and were incubated into the antibody solution in PBS (1 mg /ml) for 4 h. After rinsing the antibody-immobilized gold surfaces with PBS (pH=7.1), they were kept at 50 mM Tris buffer solution. Then, they were immersed in a solution of PGK (0.05 µM) in 50 mM tris buffer in a MAKTEK glass bottom petri-dish kept on a permanent magnet for different time intervals (2 s, 4 s, 6 s, 8 s) and immediately taken out and rinsed with buffer. The reaction kinetics were studied with both orientations (either H+ or H-) of the magnet and also in absence of the magnetic field as a control. The interaction was also studied in the presence of ADP (2.5 mM) or ATP (2.5 mM) in a magnesium chloride (3 mM) solution. Fluorescence imaging was carried out immediately following sample preparation. All samples were prepared twice to test reproducibility of the results.

*Microscopy experiments & data analysis*

The fluorescence imaging of the samples was done using a home-built total internal reflection fluorescence microscope (TIRFM). A detailed description of the TIRFM setup is given elsewhere.[2] We recorded the data and analyzed them following the same procedure as reported in the Ref 1. In each experiment, 10 different TIRFM movies were recorded on 10 different regions with a size of 101 pixel X 101 pixel, i.e. 6.73µm X 6.73 µm of the sample. On each region, we recorded 100 ms frames until all molecules in the designated area were photo bleached. TIRFM movies were analyzed using custom-written Matlab (MathWorks) routines. Individual spots were identified in the first frame of a movie using a combination of thresholding and center of mass (CM) analysis as described previously.[3]

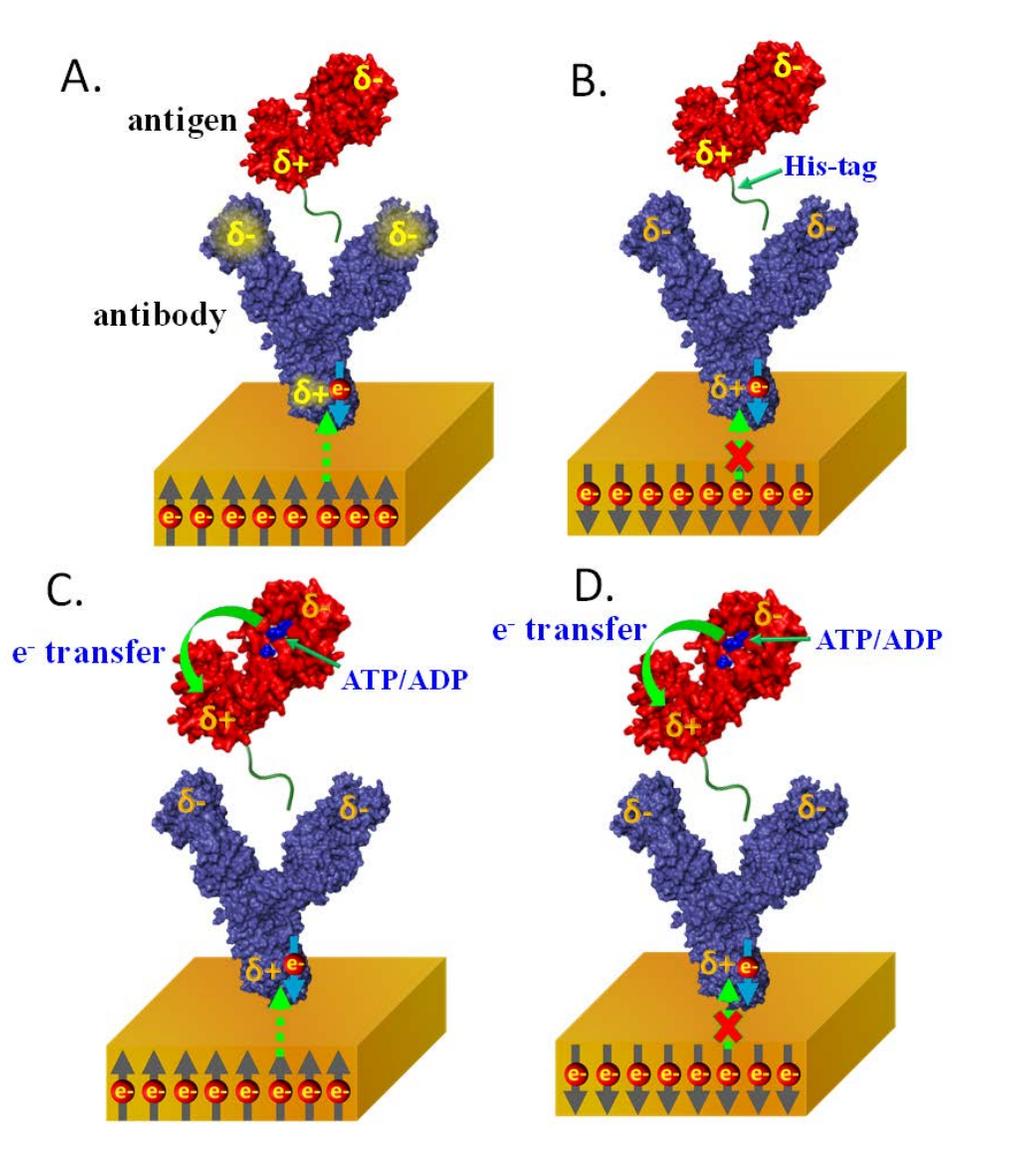

**Supporting figure 1:** Schematic representation of the role of the CISS effect in the proposed explanation for the experimental results. (A) as in our previous study, one direction of substrate magnetization facilitates electron transfer from metal substrate to the antibody following polarization of the latter by the approaching antigen; the result is a faster protein-protein association process. (B) When the substrate is magnetized in the opposite direction, charge flow into the antibody is hindered and results in an inefficient CRA in the antibody and a slow association reaction. (C-D) In the presence of ADP or ATP, changes in the polarization of the antigen lead to less positive charge close to its C terminus and therefore less electrons are attracted from the substrate, either when electron transfer is allowed (C) or not (D).